\documentclass[12pt]{article}
\usepackage[margin=1in]{geometry}% <--- 1 in margin
\usepackage{hyperref}
\usepackage{amsmath}
\usepackage{setspace}
\usepackage{graphicx}
\usepackage{xcolor}
\usepackage{authblk}
\doublespace

\title{Extreme Compactness, Extreme Gravity: Higher-Derivative Corrections to ECOs\footnote{Essay awarded honorable mention for the Gravity Research Foundation 2025 Awards for Essays on Gravitation}}
\vspace{10pt}
\author{\textbf{Madhur Mehta}\thanks{mehta.493@osu.edu (corresponding author)}  % Author in bold with email as footnote
\\
\vspace{20pt}
Department of Physics\\\vspace{-10pt}
The Ohio State University\\\vspace{-10pt}
Columbus, OH 43210, USA}
\date{}

\newcounter{MMaskc}
\begin{document}

\maketitle
\begin{abstract}
    Higher-derivative gravity theories offer insights into the behavior of extremely compact objects (ECOs). Focusing on Gauss-Bonnet (GB) and Einstein-dilaton-Gauss-Bonnet (EdGB) gravity, we derive the compactness scale in these models and demonstrate how higher-curvature corrections lead to deviations from the standard ECO compactness scale. The corrections are of order $\alpha/r_0^2$ in EGB gravity and $\alpha^2/r_0^4$ in EdGB gravity, where $\alpha$ is the coupling constant and $r_0$ is the horizon radius corresponding to the mass of the ECO. Observational constraints suggest these effects could be significant in certain astrophysical systems, providing a new perspective on the nature of extremely compact objects in models of modified gravity.
\end{abstract}
\newpage
\section{Introduction\label{sec1}}
Modified theories of gravity \cite{Lovelock:1971yv} and the nature of compact objects therein \cite{Carballo-Rubio:2025fnc} have emerged as crucial topics in modern theoretical physics, particularly in the context of quantum gravity and black hole microstructure \cite{Mathur:2024ify}. Although General Relativity (GR) offers a simple, elegant, and highly successful framework for describing astrophysical black holes, it remains incomplete at a fundamental level \cite{PhysRevD.16.953}. Among the leading candidates for these modifications are theories incorporating higher-curvature corrections to the Einstein gravity, such as those arising from string theory or effective field theory considerations. A particularly well-studied class of gravity theories is the Gauss-Bonnet (GB) gravity model, where the Einstein-Hilbert action is extended by quadratic curvature terms that naturally arise in low-energy string-theoretic corrections \cite{Moura:2006pz}. The GB term $\mathcal{R}^2_{GB}$ is significant because it is the only quadratic Lagrangian that leads to second-order field equations for the metric \cite{Lovelock:1971yv}. Notably, the GB term, is a topological invariant in four-dimensional spacetime ($D = 4$) and does not contribute to the dynamical equations of motion if included in the action. However, when coupled to a dilaton field, this term becomes dynamical, making it an interesting correction to the Einstein-Hilbert action in $D = 4$ dimensions. The Einstein-dilaton-Gauss-Bonnet (EdGB) gravity is thus obtained by introducing a real scalar field, the dilaton, which is non-minimally coupled to the GB term. Recent constraints on the coupling constant $\alpha$ in the EdGB model, derived from gravitational wave observations, suggest that $ \sqrt{\alpha} \leq 0.295$ km \cite{Gao:2024rel}. However, most other constraints indicate that \( \sqrt{\alpha} \) is typically on the order of a few kilometers. These constraints suggest that while $\alpha$ is theoretically expected to be on the order of the Planck length, based on the only length scale available in the theory, observational data allow for values up to a few kilometers, making corrections of order $\alpha$ and $\alpha^2$ significant in certain astrophysical contexts.

Lessons from the black hole information paradox indicate that quantum gravitational effects are expected to modify black hole solutions \cite{Mathur:2024ify}. This could give rise to horizonless alternatives known as Extremely Compact Objects (ECOs) \cite{Mathur:2023uoe}. Such objects, which mimic black holes but exhibit deviations in their near-horizon structure, offer a compelling arena for testing quantum gravity-inspired corrections. The presence of these corrections modifies the structure of compact objects, potentially altering their gravitational-wave signatures in a manner that can be probed observationally.

In this work, we investigate the impact of EGB and the EdGB on the compactness scale of ECOs. We show that the addition of higher-curvature terms with a coupling constant $\alpha$ leads to a correction to the compactness scale. We find that in EGB gravity ($D>4$ dimensions), the compactness scale receives order $\alpha/r_0^2$ correction while in EdGB gravity ($D=4$), the first corrections appear at order $\alpha^2/r_0^4$ where $r_0$ is the horizon radius for the mass of the ECO.

\section{String-generated gravity models\label{sec2}}
Higher-order gravity models are typically motivated by ultraviolet (UV) corrections to the classical Einstein-Hilbert action. Additionally, some of these models arise as low-energy truncations of string theories. In this work, we focus on two such models: Einstein-Gauss-Bonnet gravity and Einstein-dilaton-Gauss-Bonnet gravity specifically for $D=4$ dimensions and lay down the specifics of these models.

\subsection{Einstein-Gauss-Bonnet Gravity\label{sub2.1}}
The effective gravity theory characterized by the Einstein plus the GB term is
\begin{align}
    S_{EGB}=\int d^{d+1}x \frac{\sqrt{-g}}{16\pi G} \left(R + \frac{\alpha}{4}\mathcal{R}^2_{GB} \right)\,,
\end{align}
where the $\mathcal{R}^2_{GB}= R_{\mu\nu\rho\sigma}R^{\mu\nu\rho\sigma}-4R_{\mu\nu}R^{\mu\nu} + R^2$ is the GB term.
In \cite{PhysRevLett.55.2656}, it was found that this action admits a Schwarzschild-like solution in $D=d+1$ dimensions of the form
\begin{align}\label{eq.metric}
  ds^2 = -A(r)dt^2+ B(r)^{-1}dr^2+r^2d\Omega_{d-1}^2\,,
\end{align}
which for small coupling $\alpha$ takes the form
\begin{align}
A(r)=B(r)\approx 1 -\frac{2GM}{r^{d-2}}+ \alpha\frac{4G^2M^2}{r^{2(d-1)}}\,,
\end{align}
which takes the usual Tangherlini-Schwarzschild form \cite{tangherlini1963} for $\alpha= 0$. It is easy to see that not all values of $\alpha$ admit black hole solutions. Our focus will be on the values that give $A(r_0)=0$ for some real value of $r_0$.
\subsection{Einstein-dilaton-Gauss-Bonnet Gravity\label{sub2.2}}
The EdGB action is
\begin{align}
    S_{EdGB}=\int d^{4}x \frac{\sqrt{-g}}{16\pi G}\left(R -\frac{1}{2}\partial^\mu \phi \partial_\mu \phi + \frac{\alpha}{4}e^{\phi}\mathcal{R}^2_{GB} \right)\,,
\end{align}
where $\alpha$ is the coupling constant and $\phi$ is the dilaton.

Static spherically symmetric black holes in the EdGB theory can be written as a perturbative series in the coupling constant. In \cite{Mignemi:1992nt}, it was shown that the first contribution to the field equations occurs only at order $\alpha^2$. 
Restricting to first correction, we find the metric \eqref{eq.metric} for $d=3$ with functions
\begin{align}
    A(r) &\approx 1-\frac{2GM}{r} + \zeta^2 \left(\frac{GM}{r}\right)^3A_2(r)\,,
    \\
    B(r) &\approx 1- \frac{2GM}{r} + \zeta^2\left(\frac{GM}{r}\right)^2 B_2(r)\,,
\end{align}
for small values of the dimensionless coupling constant $\zeta = \frac{\alpha}{(G M)^2}$. The functions $A_2(r)$ and $B_2(r)$ have been found in \cite{Mignemi:1992nt,Yunes:2011we} to be
\begin{align}
    A_2(r) &= \left(1+26\frac{GM}{r}+\frac{66}{5}\frac{G^2M^2}{r^2}+\frac{96}{5}\frac{G^3M^3}{r^3}-80\frac{G^4M^4}{r^4}\right)\,,\nonumber
    \\
    B_2(r)&=\left(1+\frac{GM}{r}+\frac{52}{3}\frac{G^2M^2}{r^2}+2\frac{G^3M^3}{r^3}+\frac{16}{5}\frac{G^4M^4}{r^4}-\frac{368}{3}\frac{G^5M^5}{r^5}\right)\nonumber\,,
\end{align}
and the horizon $r_0$ is found to lie at $r_0 \approx (2-\zeta^2)GM$. 

\section{Extremely Compact Objects in modified gravity\label{sec3}}
\begin{figure}
    \centering
    \includegraphics[width=0.7\linewidth]{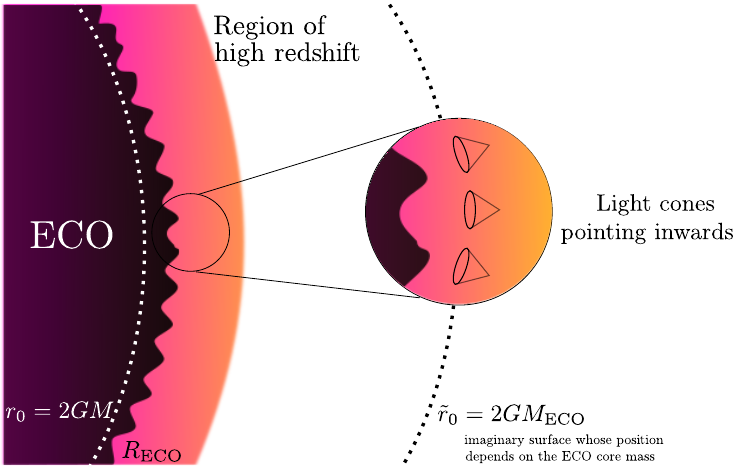}
    \caption{Schematic description of the ECO argument for an ECO with $T_{\text{ECO}}=0$ and a  surface $r=R_{\text{ECO}}$  that is just a planck length outside the horizon radius. The region just outside $R_{\text{ECO}}$ has a large negative energy density due to the negative vacuum energy. Thus the core of the ECO (the region depicted with a jagged boundary) must have a mass $M_\text{ECO}$ significantly more than $M$ and this core must be inside its own horizon. Thus such an ECO cannot exist as a time-independent configuration. This argument extends to all $T_\text{ECO}<T_\text{H}$ and a similar argument prohibits ECOs with $T_{\text{ECO}}>T_\text{H}$.}
    \label{fig:ECO1}
\end{figure}

An extremely compact object (ECO) is defined as a quantum object without horizon, whose radius is just a small proper distance $s$ outside  its Schwarzschild radius $r_0$. Fig. \ref{fig:ECO1} provides a schematic representation of the ECO argument, showing the mass distribution and the near-horizon structure. In \cite{Mathur:2024mvo}, it was shown that any ECO of mass $M$ in $d+1$ dimensions obeying the ECO conditions mentioned below, must have (at leading order) the same thermodynamic properties --- Hawking temperature $T_\text{H}$, entropy and  radiation rates  --- as the corresponding semiclassical black hole of mass $M$.

\subsection{Criteria for Identifying ECOs}
An ECO is a quantum object that exhibits the following characteristics:
\begin{enumerate}
    \item[] \textbf{ECO1:} Semiclassical physics holds outside the ECO radius $r=R_\text{ECO}$ and $R_\text{ECO}$ is close to the corresponding horizon $r_0$ with its proper radius being
        \begin{align}
            s_\text{ECO} \ll \left( \frac{M}{m_p} \right)^{\frac{2}{(d-2)(d+1)}} l_p \,\sim\, \left( \frac{r_0}{l_p} \right)^{\frac{2}{d+1}} l_p\,.
        \end{align}
    \item[]  \textbf{ECO2:} The redshift at $r=R_\text{ECO}$ satisfies
        \begin{align}
            q(R_{\text{ECO}})\gg  \left( \frac{r_0}{l_p} \right)^{\frac{d-1}{d+1}} \sim \left( \frac{M}{m_p} \right)^{\frac{d-1}{(d-2)(d+1)}}\,,
        \end{align}
        implying a large gravitational redshift at the ECO’s surface. 
    \item[] \textbf{ECO3:} At distances $s>\left( \frac{M}{m_p} \right)^{\frac{2}{(d-2)(d+1)}} l_p$ from the black hole horizon radius $r_0$, the geometry is well approximated by the corresponding black hole metric.
\end{enumerate}
We define $s_c\equiv \left( \frac{M}{m_p} \right)^{\frac{2}{(d-2)(d+1)}} l_p$ as the compactness scale for any ECO.
\subsection{Heuristic derivation of the compactness scale $s_c$}\label{sub3.1}
The compactness scale $s_c$ for an Extremely Compact Object (ECO) of mass $M$ (as measured at infinity) can be understood as the critical scale at which the total energy—comprising the ECO's mass and the vacuum energy outside it—becomes sufficient for gravitational collapse at the ECO's radius. This is illustrated in Fig.~\ref{fig:ECO1}. Mathematically, the compactness scale is determined by imposing the condition  
\begin{align}
    g_{tt}(r=R_\text{ECO},M=M_\text{ECO}) = 0\,,
    \label{eq.colcon}
\end{align}  
where $R_\text{ECO} = r_0 + \delta r$ is the ECO radius, with $r_0$ denoting the corresponding horizon and $\delta r$ representing the small coordinate separation between them. The total mass of the ECO, including the vacuum energy contribution, is given by  
\begin{align}
    M_\text{ECO} = M + E_\text{vac}\,,
\end{align}  
where $M$ is the mass observed at infinity, and $E_\text{vac}$ accounts for the energy in the vacuum surrounding the ECO.  

For a static, spherically symmetric ECO, the exterior metric component satisfies  
\begin{align}
    g_{tt}(r\geq R_\text{ECO},M) = 1-\frac{2 GM}{r^{d-2}}
\end{align}  
in $D = d+1$ spacetime dimensions, as dictated by the ECO3 condition. From the collapse condition \eqref{eq.colcon}, we can express the vacuum energy $E_\text{vac}$ in terms of $\delta r$ by substituting $R_\text{ECO} = r_0 + \delta r$ and $M_\text{ECO}= M + E_\text{vac}$ while retaining only the leading-order terms. Near the ECO surface, invoking ECO3, the coordinate distance $\delta r$ can be rewritten in terms of the proper distance (Rindler coordinate $s$) as $\delta r \propto s^2$. Consequently, the vacuum energy scales as $E_\text{vac} \propto s^2$. Additionally, one can compute the energy density and, consequently, the total energy in the vacuum within the Schwarzschild frame near the high-redshift region of the ECO, as a function of the proper distance $s$ from the horizon $r_0$:
\begin{align}
    E_\text{vac} \propto \frac{1}{s^{d-1}}\,,
    \label{eq.enprop}
\end{align}  
where $s$ is the proper distance from the horizon $r_0$.  

By combining the implication of the collapse condition \eqref{eq.colcon} ($E_\text{vac} \propto s^2$) with the scaling relation for vacuum energy \eqref{eq.enprop}, the compactness scale $s_c$ is obtained as  
\begin{align}
    s_c= \left ( {M\over m_p}\right )^{2\over (d-2)(d+1)} l_p\,,
\end{align}  
as shown in \cite{Mathur:2023uoe}.

\subsection{Corrections to the compactness scale $s_\text{c}$ in higher derivative gravity\label{sec4}}

Using the heuristic derivation for higher-order gravity models, we derive the compactness scale for ECOs in EGB gravity with the corresponding horizon at $r_0$, (see subsection \ref{sub2.1}):
\begin{equation}
    s^{EGB}_c \sim \left(\frac{\, r_0^d}{M}\left( 1-\alpha\frac{4GM}{r_0^d}\right)\right)^\frac{1}{d+1}l_p\,.
\end{equation}
As a perturbative expansion in $\alpha$, we find
\begin{align}
    s^{EGB}_c &\sim \left(\frac{M}{m_p}\right)^\frac{2}{(d+1)(d-2)}\,l_p\left(1-\frac{4}{d+1}\frac{\alpha}{(G M)^\frac{2}{d-2}}+\mathcal{O}(\alpha^2)\right)\,
    \nonumber\\
    &\sim\left(\frac{r_0}{l_p}\right)^\frac{2}{(d+1)}\,l_p\left(1-\frac{4}{d+1}\frac{\alpha}{r_0^2}+\mathcal{O}(\alpha^2)\right)\,,
\end{align}
where we have replaced $G$ by $l_p^{d-1} = m_p^{-(d-1)}$ and used $r_0 \sim (GM)^\frac{1}{d-2}$. The compactness scale for ECOs in the EdGB gravity with the corresponding horizon at $r_0$, (see subsection \ref{sub2.2}) is:
\begin{align}
s^{EdGB}_{c} \sim  \sqrt{\frac{r_0}{l_p}}\left( \frac{1-\zeta^2 \frac{G^2 M^2}{2r_0^2}}{1-\frac{4GM}{3r_0}}\right)^\frac{1}{4}\,l_p\,,
\end{align}
where we have used the fact that $A_2(r_0) \approx 1$ to set it to $1$ with $\zeta = \frac{\alpha}{(G M)^2}$ as the dimensionless coupling constant and used $r_0 \sim GM$. As a perturbative expansion in $\alpha$, we find
\begin{align}
   s^{EdGB}_{c} &\sim \sqrt{\frac{M}{m_p}}\,l_p\,\left(1-\frac{\alpha^2}{(G M)^4} +\mathcal{O}(\alpha^4)\right)\,
    \sim\sqrt{\frac{r_0}{l_p}}\,l_p\,\left(1-\frac{\alpha^2}{r_0^4} +\mathcal{O}(\alpha^4)\right)\,.
\end{align}
\section{Summary}
This study investigates the influence of Gauss-Bonnet (GB) gravity and Einstein-dilaton-Gauss-Bonnet (EdGB) gravity on the compactness of Extremely Compact Objects (ECOs). While General Relativity (GR) successfully describes astrophysical black holes, quantum gravity considerations and solutions to the information paradox suggest modifications at higher energies, potentially leading to horizonless ECOs.

We derive corrections to the compactness scale of ECOs in these theories, finding that the presence of a coupling constant $\alpha$ introduces modifications to their near-horizon structure. In EGB gravity the compactness scale receives corrections of order $\alpha/r_0^2$ while in EdGB gravity ($D=4$), the first corrections appear at order $\alpha^2/r_0^4$. Although, theoretically, the only scale available for $\alpha$ is $l_p$, observational constraints suggest that $\alpha$ can be as large as a few kilometers. This makes corrections of order $\alpha$ and $\alpha^2$ significant and relevant. This approach can be readily extended to other gravity models, such as $f(R)$ gravity, allowing for the identification of ECO equivalents within these theories. 

\section{Acknowledgments}
I would like to thank Prof. Samir D. Mathur for his valuable insights and helpful suggestions. I would also like to thank Soumangsu Chakraborty, Brandon Manley and Kaitlyn Hillery for their help and support. This work is funded by the presidential fellowship granted by the Ohio State University.
\\
\textcolor{white}{Shanti}
\bibliographystyle{utphys}
\bibliography{mmtp}
\end{document}